\documentclass[pre,aps,twocolumn,preprintnumbers,showpacs,superscriptaddress,amsmath,amssymb]{revtex4-1}

\usepackage                             {cmap} 
\usepackage     [utf8]                  {inputenc}
\usepackage     [T1]                    {fontenc}
\usepackage                             {color}
\usepackage                             {amsmath}
\usepackage                             {graphicx}
\usepackage     [english]               {babel}
\usepackage                             {hyperref}
\usepackage	[usenames,dvipsnames]	{pstricks}
\usepackage				{dcolumn}
\usepackage				{bm}

\begin{document}

\title{Stability of Quantum Statistical Ensembles with Respect to Local Measurements}

\author{Walter Hahn}
\email{w.hahn@thphys.uni-heidelberg.de}
\affiliation{Skolkovo Institute of Science and Technology, Skolkovo Innovation Centre, Nobel Street 3, Moscow 143026, Russia}
\affiliation{Institute for Theoretical Physics, Philosophenweg 19, 69120 Heidelberg, Germany}

\author{Boris V. Fine}
\email{b.fine@skoltech.ru}
\affiliation{Skolkovo Institute of Science and Technology, Skolkovo Innovation Centre, Nobel Street 3, Moscow 143026, Russia}
\affiliation{Institute for Theoretical Physics, Philosophenweg 19, 69120 Heidelberg, Germany}

\begin{abstract}
We introduce a stability criterion for quantum statistical ensembles describing macroscopic systems. An ensemble is called ``stable'' when a small number of local measurements cannot significantly modify the probability distribution of the total energy of the system. We apply this criterion to lattices of spins-1/2, thereby showing that the canonical ensemble is nearly stable, whereas statistical ensembles with much broader energy distributions are not stable. In the context of the foundations of quantum statistical physics, this result justifies the use of statistical ensembles with narrow energy distributions such as canonical or microcanonical ensembles.
\end{abstract}


\maketitle

\section{Introduction}
An isolated classical system always has a fixed value of the total energy. In contrast, an isolated quantum system can be in a superposition of states with different total energies~\cite{schrodinger}. This entails the following difficulty for the foundations of statistical physics: A typical isolated macroscopic quantum system is generally expected to thermalise under the action of its internal dynamics for the overwhelming majority of initial nonequilibrium states appearing in nature or created in a laboratory~\mbox{\cite{rigol_review,lauchli,HoLee,gemmer,goldstein_leb,popescu_short,reimann,banuls_cirac,hanggi,rigol,rigol_arxiv}}. Thermalisation implies that the density matrix of any small subsystem within the large system approaches a form consistent with the canonical Gibbs density matrix (canonical ensemble) for the large system. A necessary condition for an isolated many-particle quantum system to thermalise is a sufficiently narrow initial probability distribution of its total energy, as is, for example, the case for the canonical and the microcanonical statistical ensembles (see below). At the same time, for an isolated quantum system, the probability of occupying an energy eigenstate remains unchanged with time. Thus, a large isolated quantum system with a broad initial distribution of the total energy thermalises not to a conventional equilibrium state with a well-defined temperature but rather to a mixture or superposition of states with different temperatures. The problem now is that the initial states characterised by the narrow distribution of total energy are {\it not} necessarily the most probable ones. For example, non-Gibbs equilibrium for small subsystems emerges when initial states are selected in the Hilbert space of a typical many-particle system among quantum superpositions with a given energy expectation value and without any constraint on the width of the energy window for participating eigenstates~\mbox{\cite{qmc,fresch_i,fresch_ii,hantschel}}. The latter condition defines the ``quantum micro-canonical'' (QMC) ensemble~\cite{wooters,brody_hughston,aarts_wetterich,fresch,qmc,eisert}.

Given the above considerations, why do the initial nonequilibrium quantum states of macroscopic systems not normally exhibit the broad participation of energy eigenstates and hence non-Gibbs statistics for small subsystems? In this paper, we address the above question by introducing the criterion of stability of quantum statistical ensembles with respect to local measurements and then apply this criterion to lattices of spins 1/2. Similar ideas in other contexts have been considered in Refs.~\cite{shimizu,donker}.

A quantum statistical ensemble  for a macroscopic system is defined by the probability $p(E)$ of occupying an eigenstate of total energy $E$. Given the density of energy states $\nu(E)$, the probability distribution of the total energy is
\begin{equation}
g(E) =  p(E) \nu(E).
\end{equation}
We call $g(E)$ broad, when $w_g/(E_\text{av}-E_{\min})\sim1$, where $w_g^2$ is the variance of $g(E)$, $E_\text{av}$ is the average energy, and $E_{\min}$ is the ground-state energy of the system. As shown in Appendix~\ref{app_width}, a canonical ensemble with a positive temperature $T$ is narrow from the above perspective, because, in this case, $w_g/(E_\text{av}-E_{\min})\sim 1/ \sqrt{N_s} \ll 1$, where $N_s\sim10^{23}$ is the number of particles or microscopic subsystems in the system, cf. Ref.~\cite{hantschel}.

The article is organised as follows. Section~\ref{sec_stab} provides the definition of the stability criterion for quantum statistical ensembles. In Sec.~\ref{sec_effect}, the effects of local measurements are discussed qualitatively. Section~\ref{sec_spinlat} includes an analytical investigation of the stability of quantum ensembles for lattices of spins-1/2, including noninteracting spins in a magnetic field and systems of interacting spins. Section~\ref{sec_numerical} presents the results of numerical investigations for systems of interacting spins. Final remarks and conclusions are given in Sec.~\ref{sec_conclude}. Longer derivations are included in the Appendices.

\section{Definition of the stability criterion} \label{sec_stab}
Let us now observe that accidental measurements of microscopic particles in a macroscopic system cannot be excluded under any foreseeable natural or experimental conditions. We, therefore, introduce the following stability criterion: {\it A physically realisable quantum statistical ensemble describing a stationary state of a macroscopic system must be stable with respect to a small number of any arbitrarily chosen local measurements within the system. The measurement is called ``local'' if the measured quantity is localised in the three-dimensional physical space~\footnote{In principle, the criterion allows for a broader definition of local measurements in the sense that linear combinations of local measurements can also be admitted.}. The number of measurements $n$ is called small if $n \ll \sqrt{N_s}$. The ensemble is called stable if}
\begin{equation} \label{DG}
\Delta G(n) \equiv \int_{- \infty}^{+ \infty} \left| g_n(E) - g_0(E) \right| dE \ll 1,
\end{equation}
{\it where $g_0(E)$ and $g_n(E)$ are the probability distributions before and after the measurements, respectively.}

The measurements in question are implied to occur naturally, for example, when a passing photon becomes entangled with the system and accidentally measured later. Without the measurement, the above process would describe a decoherence event~\cite{zurek_rmp,joos_zeh_1985,joos_zeh_book,Leggett_macro_2,katsnelson_spinbath,schlosshauer_book}.

In the following, we simplify the analysis by assuming random instantaneous projective measurements of individual particles in the system~\cite{gardiner_book,ballesteros,guerlin_haroche,Bauer_Bernard,maassen_kumm}. Furthermore, we mainly focus on ``strong instability'', which we define as the case when fewer than 10 measurements lead to $\Delta G\gtrsim0.1$ independent of $N_s$.

\section{Effects of local measurements} \label{sec_effect}

\subsection{Narrowing vs. broadening}
Qualitatively, measurements can lead to both narrowing and broadening of $g(E)$. The broadening effect of a single measurement is due to the off-diagonal elements of the projection operator describing the measurement in the basis of the total-energy eigenstates. The narrowing effect originates from correlations between the total energy of the system and the measurement outcomes. Indeed, broad probability distributions $g(E)$ can be considered as a mixture of microcanonical (or canonical) ensembles corresponding to different temperatures $T(E)$, which, in turn, imply different probability distributions of local variables. When a given measurement outcome is more likely for one represented temperature than for another, the post-measurement distribution will be narrower than the initial one~\cite{ballesteros}. In terms of the energy scales, the increase of the variance $w_g^2$ due to the off-diagonal elements of a local projection operator should normally be of the order of $\epsilon_1^2$, where $\epsilon_1$ is an appropriately chosen single-particle energy. The reason is that the quantum projections corresponding to the local measurements are one-particle or a few particle operators. At the same time, the decrease of $w_g^2$ for broad $g(E)$ can easily be of the order $\epsilon_1^2N_s^2$, i.e., much larger. Since we focus primarily on broad $g(E)$, we neglect the broadening effect of the measurements unless explicitly specified otherwise. We further limit our derivations to $g(E)$ satisfying inequality
\begin{equation} \label{eqn_ineq}
 \left|\frac{dg(E)}{dE}\right|\lesssim \frac{g(E)}{w_\text{can}},
\end{equation}
where $w_\text{can}$ is the width of the energy distribution corresponding to the canonical ensemble with the same average energy as that of $g(E)$.

\subsection{Heating effect of measurements}
For an ensemble with narrow energy distribution $g(E)$, such as the canonical ensemble, the above-mentioned broadening effect also leads to heating, defined as the drift of the average energy $E_\text{av}$ in the direction of larger entropy $S(E_\text{av})$ (larger value of $\ln[\nu(E_\text{av})]$). This means that $E_\text{av}$ increases for positive temperatures and decreases for negative ones. Such a behaviour is consistent with the second law of thermodynamics, because local measurements of individual particles can be viewed as small nonadiabatic perturbations of the system. For positive temperatures $T$, the increase of $E_\text{av}$ due to one measurement is of the order of the one-particle energy $\epsilon_1$.

A broad ensemble can be considered as a mixture of canonical ensembles with different temperatures. The heating for each of the contributing canonical ensembles occurs as explained above. Therefore, the overall heating effect for a broad ensemble corresponds to the combination of the heating effects for the individual canonical ensembles. In a system with a finite Hilbert space, the asymptotic shape of $g(E)$ corresponds to the canonical ensemble at infinite temperature, which is, in turn, proportional to the density of states $\nu(E)$ of the system.

\subsection{Coarse-graining of the energy axis} \label{sec_coarse}
Our definition of $g(E)$ implies averaging over energy bins whose width $\Delta_{\text{e}}$ satisfies the following inequalities:
\begin{equation} \label{bins}
\epsilon_1  \ll \Delta_{\text{e}}  \ll T(E_\text{av}) \sqrt{C_V(E_\text{av})}.
\end{equation}
In the case of negative temperatures, $|T(E_\text{av})|$ should be used instead.

The left inequality in Eq.~\eqref{bins} together with the restriction $n\ll\sqrt{N_s}$ and the inequality~\eqref{eqn_ineq} allows us to neglect heating described in the preceding part. The right inequality in Eq.~\eqref{bins} assures that the energy eigenstates within each bin approximately correspond to the same density matrices of small subsystems within the entire system considered.

The probability distribution $g_n(E)$ is defined in terms of the above energy bins as follows
\begin{equation} \label{eqn_defg}
 g_n(E)=\frac{1}{\Delta_\text{e}}\sum_k^{\text{bin}(E)}(\rho_n)_{kk},
\end{equation}
where the sum is taken over all energy eigenstates within the given bin and $(\rho_n)_{kk}$ are the diagonal elements of the density matrix $\rho_n$ represented in the basis of the total-energy eigenstates.

\section{lattices of spins-1/2} \label{sec_spinlat}
Now we consider a lattice of $N_s$ spins-1/2 and examine how multiple random local measurements affect $g(E)$. We implement an individual local measurement by selecting a random spin at a random time and then measuring its projection on a random axis. The measurements are assumed to occur very rarely with constant average rate per spin $\tau_\text{m}^{-1}$ ($\tau_\text{m}$ is much longer than the characteristic time of microscopic dynamics). We label measurements by index $n$. Each measurement is characterised by the parameters $\{ m_n, \vartheta_n, \varphi_n  \}$, where $m_n$ labels the lattice site of the measured spin and $(\vartheta_n, \varphi_n)$ are the polar and azimuthal spherical angles  indicating the orientation of the spin after the measurement.

\subsection{Projection operator}
The projection operator ${\cal P}_n$, which represents the $n$th measurement, is defined as
\begin{equation} \label{eqnrojector}
{\cal P}_n\equiv\cdots\mathbf{1}_{m_n-1}\otimes\left(|\vartheta_{n}\varphi_{n}\rangle\langle\vartheta_{n}\varphi_{n}|\right)_{m_n}\otimes\mathbf{1}_{m_n+1}\cdots,
\end{equation}
where $\mathbf{1}_i$ is the unit matrix acting on the Hilbert space associated with the spin at lattice site $i$ and $|\vartheta_{n}\varphi_{n}\rangle=\cos\left(\frac{\vartheta_n}{2}\right)|\!\!\uparrow\rangle_z+\sin\left(\frac{\vartheta_n}{2}\right)e^{i\varphi_n}|\!\!\downarrow\rangle_z$ is the quantum state of a spin polarised into the direction given by the spherical angles $(\vartheta_{n}, \varphi_{n})$. The operator $\left(|\vartheta_{n}\varphi_{n}\rangle\langle\vartheta_{n}\varphi_{n}|\right)_{m_n}$ acts on the Hilbert space associated with the spin at lattice site $m_n$. The operator ${\cal P}_n$ satisfies the relations ${\cal P}_n^\dagger={\cal P}_n$ and ${\cal P}_n{\cal P}_n={\cal P}_n$.

The operator ${\cal P}_n$ is related to the operator $S_n$ of the $m_n$th spin projection in the direction $(\vartheta_{n}, \varphi_{n})$~\cite{sakurai} as
\begin{equation} \label{eqn_sp}
{\cal P}_n= \frac{1}{2}\mathbf{1} + S_n,
\end{equation}
where  $\hbar=1$.

\subsection{Evolution of the energy distribution $g(E)$: general formalism}
We denote the density matrix of the total system after $n$ measurements as $\rho_n$ and assume that the initial density matrix $\rho_0$ is diagonal in the basis of the energy eigenstates (see Appendix~\ref{app_derivation_5}). The transformation from $\rho_{n-1}$ to \mbox{$\rho_n$ reads}
\begin{equation} \label{eqn_rho_next}
 \rho_n=\frac{{\cal P}_ne^{-i{\cal H}(t_{n}-t_{n-1})}\rho_{n-1}e^{i{\cal H}(t_{n}-t_{n-1})}{\cal P}_n^\dagger}{\text{Tr}\left[{\cal P}_ne^{-i{\cal H}(t_{n}-t_{n-1})}\rho_{n-1}e^{i{\cal H}(t_{n}-t_{n-1})}{\cal P}_n^\dagger \right]},
\end{equation}
where $t_{n}$ is the time of the $n$th measurement. The corresponding probability distribution of the total energy after $n$ measurements $g_n(E)$ can be expressed as
\begin{eqnarray} \label{eqn_gen}
 g_n(E)=\frac{1}{B}\Big[&& {\cal P}^{\dagger}_1e^{i{\cal H}(t_{2}-t_{1})}{\cal P}^{\dagger}_{2}\cdots{\cal P}^{\dagger}_{n} \\
 && \times {\cal P}_{n}\cdots{\cal P}_{2}e^{-i{\cal H}(t_{2}-t_{1})}{\cal P}_1 \Big]_{\text{diag}}\!(E)\ g_0(E),\nonumber
\end{eqnarray}
where $B$ is a normalisation factor, and $[ \cdots ]_{\text{diag}}\!(E)$ denotes the diagonal elements of the operator in the energy basis averaged over suitably chosen energy bins introduced in Sec.~\ref{sec_coarse}. The derivation of Eq.~\eqref{eqn_gen} is given in Appendix~\ref{app_derivation_5}.

Since the measured spins are typically far away from each other, the effect of individual measurements in Eq.~\eqref{eqn_gen} normally factorises, which leads to
\begin{equation} \label{eqn_av_app}
 g_n(E)=\frac{1}{B_n} \left[{\cal P}_n\right]_{\text{diag}}\!(E)\ g_{n-1}(E),
\end{equation}
where $B_n$ is a normalisation factor. The derivation of Eq.~\eqref{eqn_av_app} is given in Appendix~\ref{app_derivation_6}. Below we consider the case when two measured spins are accidentally close to each other such that the corresponding effect does not factorise.

The stability measure~\eqref{DG} averaged over all possible outcomes of $n$ measurements reads (see Appendix~\ref{app_anexp})
\begin{equation} \label{eqn_dg_av}
 \overline{\Delta G}(n)\!=\overline{\int\bigg|\prod_{i=1}^n\frac{1}{B_i}\left[ {\cal P}_i \right]_{\text{diag}}\!(E) - 1\bigg|g_0(E)dE},
\end{equation}
where the bar denotes the result of averaging.

\subsection{Noninteracting spins in a magnetic field}
Let us now turn to the example of noninteracting spins in magnetic field $H_z$ with Hamiltonian
\begin{equation} \label{eqn_ham1}
 {\cal H}=- H_z\sum_iS_{iz},
\end{equation}
where ${S}_{ix}$, ${S}_{iy}$, and ${S}_{iz}$ are the spin operators on the \mbox{$i$th} lattice site. In this case, the outcome of a single spin measurement normally correlates with the total energy of the system and, therefore, leads to a significant narrowing of $g(E)$ governed by Eq.~\eqref{eqn_av_app}. A calculation based on Eq.~\eqref{eqn_sp} gives (see Appendix~\ref{app_derivation_9})
\begin{equation} \label{eqn_enmod_h}
 \left[{\cal P}_n\right]_{\text{diag}}\!(E) = \frac{1}{2}-\cos(\vartheta_{n})\frac{E}{E_{\max}-E_{\min}},
\end{equation}
where $E_{\max}= H_zN_s/2$ and $E_{\min}=-H_zN_s/2$.
\begin{figure}[b]
	\centering
	\includegraphics[width=0.48\textwidth]{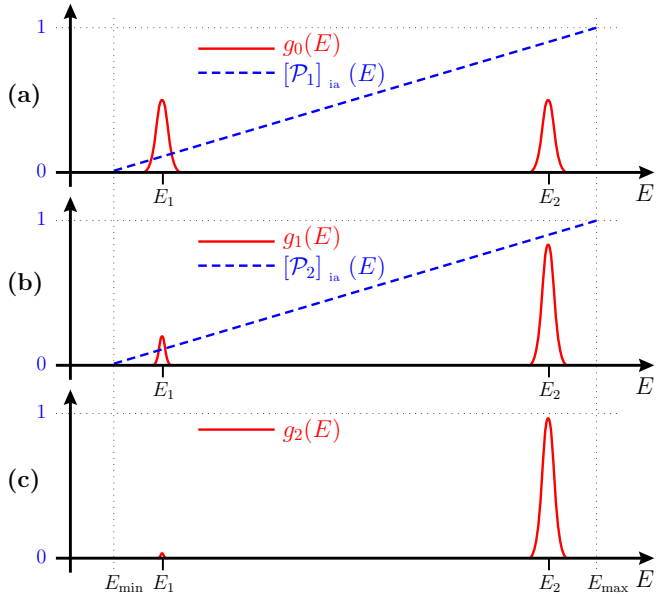}
	\caption{(Colour online) Schematic representation of the evolution of a two-peak energy distribution $g_n(E)$ (solid red lines) governed by Eq.~\eqref{eqn_av_app}: $g_0(E) = 1/2[\delta(E - E_1) + \delta(E - E_2)]$; $g_1(E) \cong  \left[{\cal P}_1\right]_{\text{diag}}\!(E)\ g_0(E)$; and $g_2(E) \cong  \left[{\cal P}_2\right]_{\text{diag}}\!(E)\ g_1(E)$. Here $\left[{\cal P}_1\right]_{\text{diag}}\!(E)$ and $\left[{\cal P}_2\right]_{\text{diag}}\!(E)$ (dashed blue lines) correspond to two single-spin measurements with respective outcomes \mbox{$\vartheta_1=\pi$} and $\vartheta_2=\pi$ substituted in Eq.~\eqref{eqn_enmod_h}.}
	\label{fig_cut}
\end{figure}

The action of transformation \eqref{eqn_av_app} consists of ``cutting'' $g_{n-1}(E)$ by function $\left[{\cal P}_n\right]_{\text{diag}}\!(E)$ and then renormalising the result. This ``cutting'' normally makes $g_n(E)$ narrower than $g_{n-1}(E)$. The outcome of the next measurement can, in principle, lead to the opposite effect, but it is more probable that it will lead to further narrowing, because the probability of subsequent measurement outcomes is determined by the narrower $g_n(E)$.  After many iterations, the drastic narrowing of $g(E)$ becomes overwhelmingly probable. 

\begin{figure}[t]
	\centering
	\includegraphics[width=0.48\textwidth]{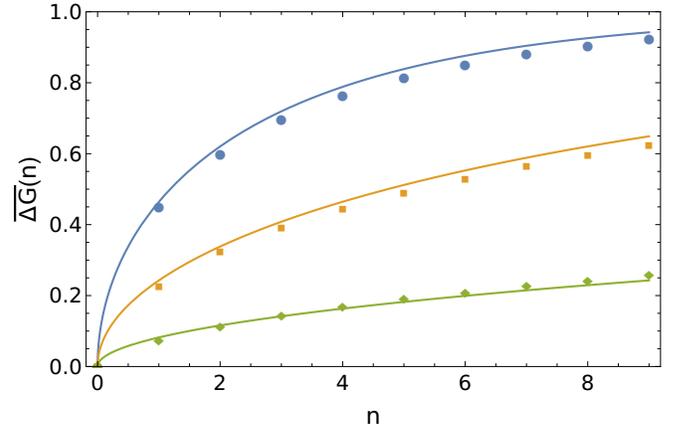}
	\caption{(Colour online) Averaged ensemble stability measure $\overline{\Delta G}(n)$ as a function of the number of measurements $n$ for a two-peak initial distribution $g_n(E)$. Points represent exact numerically computed results as explained in Appendix~\ref{app_fig}. Lines correspond to the approximated expression~\eqref{eqn_appr} with $\lambda=0.3\ \frac{(E_1-E_2)^2}{(E_{\max}-E_{\min})^2}$. Different colours represent different pairs of values for ($E_1$,$E_2$): blue (circles) ($-0.9$,$0.9$), yellow (squares) ($-0.9$,$0.0$), and green (rhombi) ($-0.9$,$-0.6$) in units where $E_{\min}\!=\!-1$ and \mbox{$E_{\max}\!=\!1$}.}
	\label{fig_stab}
\end{figure}

Figure~\ref{fig_cut} schematically illustrates the effect of a sequence of transformations \eqref{eqn_av_app} applied to the initial two-peak distribution
\begin{equation}
 g_0(E) \approx\frac{1}{2}\big[\delta(E-E_1)+\delta(E-E_2)\big],
\end{equation}
where $\delta(...)$ is a Dirac $\delta$ function~\footnote{The delta-function is an approximation for a peak with a width which is much smaller than the distance between the two peaks but still large enough to satisfy the assumptions behind the derivations of Eqs.~\eqref{eqn_gen} and~\eqref{eqn_av_app}}. In this case, one of the two peaks dominates $g_n(E)$ for $n \to \infty$. Figure~\ref{fig_stab} presents computed $\overline{\Delta G}(n)$ for the above $g_0(E)$. In each case, we obtain that, after nine measurements, \mbox{$\overline{\Delta G}(n) >0.1$} independent of the number of spins in the system, which implies ``strong instability''. In Appendix~\ref{app_anexp}, we obtained the analytical approximation
\begin{equation} \label{eqn_appr}
 \overline{\Delta G}(n) \approx \sqrt{1-e^{-\lambda n}},
\end{equation}
where $\lambda \cong u^2 (E_2-E_1)^2$ with $u \equiv  \overline{|d\left[{\cal P}_n\right]_{\text{diag}}\!(E)/dE|} \sim 1/(E_{\max}-E_{\min})$. This approximation is illustrated in Fig.~\ref{fig_stab}.

Let us now consider the initial Gaussian distribution
\begin{equation}
 g_0(E) \cong \exp{\left[-\frac{(E-E_0)^2}{2w_{g,0}^2} \right]}
\end{equation}
defined by parameters $E_0$ and $w_{g,0}$, where \mbox{$w_{g,0}\ll E_{\max} - E_{\min}$}. After $n$ measurements, $g_n(E)$ remains approximately Gaussian with the width $w_{g,n}$ following the relation
\begin{equation}
 \overline{\frac{1}{w^2_{g,n}}}=\frac{1}{w_{g,0}^2}+u^2 n,
\end{equation}
which is derived in Appendix~\ref{app_gauss}. This relation leads to \mbox{$\overline{\Delta G}(n)\sim1$} when $\overline{1/w^2_{g,n}} \sim 2/w_{g,0}^2$, which corresponds to the number of measurements $n_{cr} \sim 1/(w_{g,0}^2 u^2)$. According to our criterion, the border case for the ensemble stability corresponds to $n_{cr}\sim\sqrt{N_s}$, which implies that the ensemble is unstable for $w_{g,0} \gg (E_{\max}-E_{\min})/\sqrt[4]{N_s}$. An ensemble with $w_{g,0} \lesssim (E_{\max}-E_{\min})/\sqrt[4]{N_s}$ may still become narrower due to measurements, but the criterion calls it ``stable'', because the narrowing is relatively slow.

\subsubsection*{Absolute stability of the canonical ensemble}
We finally note that, for $w_{g,0} \sim \epsilon_1\sqrt{N_s}$, where \mbox{$\epsilon_1\equiv(E_{\max}-E_{\min})/N_s$} is the characteristic single-spin energy, the decrease of the variance as a result of one measurement is $w^2_{g,1} - w_{g,0}^2 \approx w_{g,0}^4 u^2 \sim \epsilon_1^2$. This is of the same order of magnitude as the earlier mentioned increase of $w_g^2$ associated with the broadening effect of $g(E)$ caused by the off-diagonal elements of the projection operator. Therefore, it is reasonable to expect that, for some $w_{g,0} \sim \epsilon_1\sqrt{N_s}$, the narrowing effect of measurements would compensate the broadening effect, and hence such an ensemble is absolutely stable with respect to measurements. Remarkably, this $w_{g,0}$ is of the order of the width of the canonical ensemble for $T\gtrsim\epsilon_1$~\cite{marzolino,kliesch}.

\subsubsection*{Characteristic ensemble-narrowing time}
Given that the system is measured on average once per time $\tau_\text{m}/N_s$, the above estimates for $\lambda$ and for $n_{cr}$ imply that the characteristic time to gain $\Delta G\sim1$ for a broad ensemble with the initial variance $w_{g,0}^2$ is
\begin{equation}
\tau_\text{c}\sim\tau_\text{m}\frac{\epsilon_1^2N_s}{w_{g,0}^2}.
\label{tau1}
\end{equation}
For a macroscopic system, $\tau_\text{c}$ is, therefore, extremely short unless $w_{g,0} \leq \epsilon_1\sqrt{N_s}$.

\subsection{Interacting spins}
Now we turn to the Hamiltonian of nearest-neighbour interaction:
\begin{equation} \label{eqn_ham2}
 {\cal H}=-\sum_{i<j}J_x{S}_{ix}{S}_{jx}+J_y{S}_{iy}{S}_{jy}+J_z{S}_{iz}{S}_{jz},
\end{equation}
where $J_x$, $J_y$, and $J_z$ are the coupling constants. In contrast to the previous case~\eqref{eqn_ham1}, the outcome of a single-spin measurement here is not correlated with the total energy of the system, i.e., $\left[{\cal P}_n\right]_{\text{diag}}\!(E)=const.$, and, hence, does not induce narrowing of $g(E)$. At least two accidental measurements sufficiently close in space and time are required for this. Let us consider two such measurements $n$ and $n-1$ at times $t_n>t_{n-1}$. The same treatment that led to Eq.~\eqref{eqn_av_app} now gives (see Appendix~\ref{app_derivation_13})
\begin{equation}
 g_n(E)=\frac{1}{B_n}\left[{\cal A}_{n,n-1}^\dagger{\cal A}_{n,n-1}\right]_{\text{diag}}\!(E)\ g_{n-2}(E),
\end{equation}
where ${\cal A}_{n,n-1}={\cal P}_ne^{-i{\cal H}(t_{n}-t_{n-1})}{\cal P}_{n-1}$. Substituting Eq.~\eqref{eqn_sp}, we obtain in Appendix~\ref{app_derivation_13}
\begin{eqnarray} \label{eqn_modi}
 &&\left[{\cal A}_{n,n-1}^\dagger{\cal A}_{n,n-1}\right]_{\text{diag}}\!(E)\\
 &&=\frac{1}{4}+\frac{1}{2}\Big[\left\{S_{n-1}(t_{n-1}),S_{n}(t_{n}) \right\}\Big]_{\text{diag}}\!(E)\nonumber\\
 &&+\Big[S_{n-1}(t_{n-1})S_{n}(t_{n})S_{n-1}(t_{n-1})\Big]_{\text{diag}}\!(E),\nonumber
\end{eqnarray}
where $\big\{\cdots,\cdots\big\}$ is the anticommutator.

\subsubsection*{Relation to equilibrium spin correlation functions}
Once the spin orientations obtained in the \mbox{$(n-1)$-st} and the $n$th measurements are specified and the energy $E$ is fixed, the last two terms on the right-hand side of Eq.~\eqref{eqn_modi} can be considered equilibrium spin correlation functions for the microcanonical ensemble corresponding to energy $E$~\footnote{Spin correlation functions entering Eq.~\eqref{eqn_modi} are defined as $\left<{S}_{i\mu}{S}_{j\nu}\right>$, where $\left<\cdots\right>$ represents the microcanonical average. This definition is to be contrasted with the correlation functions for spin fluctuations $\left<{S}_{i\mu}{S}_{j\nu}\right>-\left<{S}_{i\mu}\right>\left<{S}_{j\nu}\right>$. The difference is essential in the presence of magnetic order.}. For macroscopic systems, these correlation functions equal, in turn, the correlation functions for the canonical ensemble with temperature $T(E)$.

To give an example, let us assume that the outcome of the first measurement is $\vartheta_{1}=0$, $\varphi_{1}=0$ (spin 1 points into the positive $z$-direction) and the outcome of the second measurement is $\vartheta_{2}=\frac{\pi}{2}$, $\varphi_{2}=0$ (spin $2$ points into the positive $x$ direction). For the two-spin term, we then obtain
\begin{equation}
 \Big[\{S_{2}(t_{2}),S_{1}(t_{1})\}\Big]_{\text{diag}}\!(E)
 \!=\!\left\langle\{ S_{x}(\vec{r}_2,t_{2}),S_{z}(\vec{r}_{1},t_{1})\}\right\rangle_{T(E)},
\end{equation}
where $\vec{r}_n$ is the position of the $n$th measured spin.

\subsubsection*{Characteristic ensemble-narrowing time}
The above spin correlation functions depend on the time delay and the distance between the two measurements. Therefore, the cutting function in Eq.~\eqref{eqn_modi} is strongly influenced by the presence of long-range magnetic order and integrals of motion. We characterise the overall behaviour of the spin correlation functions by correlation time $\tau_\text{corr}(E)$ and correlation length $\xi(E)$. If a system has constants of motion associated with spin polarisations, this corresponds to an infinite $\tau_\text{corr}(E)$. Likewise, if a system exhibits long-range magnetic ordering in a range of energies $E$, this means that for this range of energies $\xi(E)$ is infinite. [It is worth noting that the correlation functions entering Eq.~\eqref{eqn_modi} differ from the conventional correlation functions of spin fluctuations: For the latter the product of single-spin expectation values is subtracted.]

\textit{Ordered phase:} If a significant part of a broad $g_0(E)$ corresponds to temperatures within the magnetically ordered phase, then the characteristic time $\tau_\text{c}$ to gain $\Delta G \sim 1$ can be estimated by expression~\eqref{tau1}. The reason is that, in the magnetically ordered phase, $\xi(E)$ is infinite and, therefore, each measurement correlates with all previous ones in the sense that, for all subsequent measurements $k$, $[S_n(t_n)S_k(t_k)]_{\text{diag}}\!(E)\neq0$. The overall situation resembles the case of Hamiltonian~\eqref{eqn_ham1}, with the external magnetic field replaced by the local field created by the ordered neighbours of each spin.

\textit{Nonordered phase:} In the nonordered (paramagnetic) phase, $\xi(E)$ is of the order of the nearest-neighbour distance (except for the energy range in the proximity to the magnetic phase transitions). We make the assumption justified by the final result that $\tau_\text{c} \ll \tau_\text{corr}(E)$ for all energies $E$, where $\tau_\text{corr}(E)$ is the correlation time introduced above. Therefore, we set the time delay entering the cutting function in Eq.~\eqref{eqn_modi} effectively to zero. Since $\tau_\text{c}\ll \tau_\text{corr}$, the outcome of the \mbox{$n$th} measurement addressing a nearest-neighbour of a previously measured spin is correlated with the total energy of the system. The probability that the \mbox{$n$th} measurement does not address a nearest-neighbour of a previously measured spin is $P_n=1-(n-1)\frac{N_\text{NN}}{N_s}$ for $n\ll N_s$, where $N_\text{NN}$ is the number of nearest neighbours. The probability that, among $n$ measurements, there is no nearest-neighbour pair measured, is
\begin{eqnarray}
 P(n)=\prod_{k=1}^n P_n&=&\prod_{k=1}^n\left(1-(k-1)\frac{N_\text{NN}}{N_s} \right)\\
 &\approx& \exp\left(-\frac{N_\text{NN}}{N_s}\sum_{k=1}^nk\right)\approx e^{-n^2\frac{N_\text{NN}}{N_s}}\nonumber
\end{eqnarray}
for $1\ll n\ll N_s$. Using the relation $n=\frac{N_st}{\tau_m}$, we finally obtain the probability that, after time $t$, no pair of nearest neighbours is measured,
\begin{equation}
 P(t)\approx e^{-N_\text{NN}N_s\frac{t^2}{\tau_m^2}}.
\end{equation}
Accordingly, $1-P(t)$ is the probability to measure at least one pair of nearest neighbours. Therefore, for the paramagnetic phase and $w_{g,0} \sim E_{\max}-E_{\min}$, we obtain $\tau_\text{c} \sim\tau_\text{m}/\sqrt{N_s}$~\footnote{Since $\sqrt{N_s}$ is very large, $\tau_\text{c}$ is certainly much smaller than any reasonable correlation time $\tau_\text{corr}$. This is a consistency check of our initial assumption $\tau_\text{c}\ll\tau_\text{corr}$.}. If only single-spin measurements were allowed, then the latter ensemble would be called stable according to our criterion.
However, the criterion admits any local measurements, including, for example, the measurements of the total spin of two neighbouring spins. For such measurements, the estimate ~\eqref{tau1} remains valid, thus rendering the ensemble unstable.

We additionally note here that applying an external magnetic field to the paramagnetic phase would drastically shorten $\tau_\text{c}$, because, in this case, single-spin measurements described by Eq.~\eqref{eqn_enmod_h} would cause a much faster ensemble narrowing.

\subsection{General case}
Beyond the Hamiltonians~\eqref{eqn_ham1} and~\eqref{eqn_ham2}, we expect that, for a broad class of systems with short-range interactions,
local measurements whose possible outcomes correlate with the total energy of the system have a narrowing effect comparable to that of single-spin measurements for Hamiltonian~\eqref{eqn_ham1}. In the general case, the analogue of the cutting function
$\left[{\cal P}_n\right]_{\text{diag}}\!(E)$ is much more difficult to calculate directly but,
otherwise, can be reasonably expected to be nonlinear with the energy-dependent slope of the order of $1/(\epsilon_1 N_s)$. The latter implies that the estimate~\eqref{tau1} for the ensemble-narrowing time and the argument for the proximity of the canonical ensemble to the absolute stability threshold remain valid.

\begin{figure}[t]
 \centering
 \includegraphics[width=0.48\textwidth]{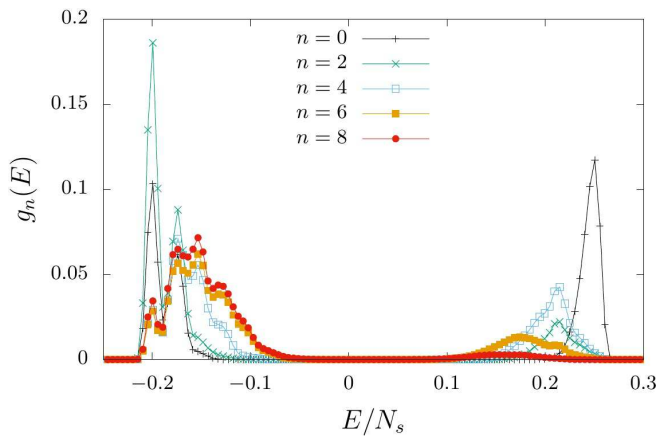}
 \caption{(Colour online) The energy distribution $g_n(E)$ is shown after each even-numbered measurement as indicated in the legend. The system size is $N_s=24$.}
 \label{fig_bins}
\end{figure}

\section{Numerical Investigation of spin systems} \label{sec_numerical}
In order to substantiate the conjectures made in the previous section, we numerically investigate a periodic chain of 24 \mbox{spins-1/2} governed by the Hamiltonian~\eqref{eqn_ham2} with \mbox{$J_x=0.47$}, \mbox{$J_y=-0.37$}, and $J_z=-0.79$. The spins are measured in pairs: An odd-numbered measurement is done on a randomly selected spin, while one of its nearest neighbours is chosen for the next even-numbered measurement. The time delays $t_n-t_{n+1}$ are chosen randomly from the interval $[0,2]$. The numerical techniques used for the calculations are described in Appendix~\ref{app_num}.

In Fig.~\ref{fig_bins}, we show a typical evolution of $g(E)$ starting with a two-peak $g_0(E)$ corresponding to the superposition of two quantum states representing canonical ensembles with temperatures $T_1=0.1$ and $T_2=-0.1$. In this example, one peak of $g(E)$ becomes significantly suppressed, which represents the narrowing sketched in Fig.~\ref{fig_cut}.

Adapting estimate~\eqref{tau1} to the present case of spin-pair measurements, we substitute $N_s = 24$, pair measurement time $\tau_m  \approx 2 N_s$,  pair energy $\epsilon_1^2 \approx 0.25 (J_x^2 + J_y^2 + J_z^2) = 0.25$, and $w_g \approx 0.2 N_s$ (see Fig.~\ref{fig_bins})  to obtain from Eq.~\eqref{tau1} $\tau_c \sim 12$ corresponding to $n \sim 12$ and within a factor of two consistent with the characteristic value $n \sim 6$ extracted from Fig.~\ref{fig_bins}.

We further tested the finite-size scaling of heating in small spin systems. As expected, the individual peaks in Fig.~\ref{fig_bins} also exhibit significant finite-size effects, namely peak broadening and the drift of the peak maxima associated with heating, which prevent us from computing $\Delta G(n)$ representative of the macroscopic limit. However, as we show below, the finite-size scaling of the heating is consistent with our expectation that they can be neglected in the macroscopic limit.

\begin{figure}[t]
 \centering
 \includegraphics[width=0.48\textwidth]{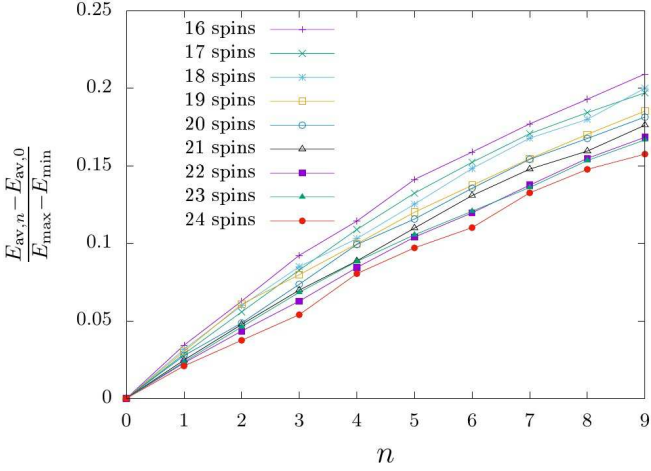}
 \caption{(Colour online) The quantity $\frac{E_{\text{av},n}-E_{\text{av},0}}{E_{\max}-E_{\min}}$ is shown for the interacting spin system introduced in Sec.~\ref{sec_numerical}. The energy distribution corresponds to the Gibbs distribution with the temperature $T=0.1$. Different system sizes were chosen as indicated in the legend. The plotted values have been averaged over many iterations. The points are connected by lines in order to guide the eye.}
 \label{fig_en_drift}
\end{figure}

In order to quantify the heating for the small system sizes available in numerical investigations, we calculate the deviation of the average energy $E_\text{av}$ from its initial value $E_{\text{av},0}$ for a one-peak $g(E)$ corresponding to the canonical ensemble with the temperature $T=0.1$. In Fig.~\ref{fig_en_drift}, we show the evolution of $\frac{E_{\text{av},n}-E_{\text{av},0}}{E_{\max}-E_{\min}}$ as a function of $n$. In this case, we randomly measure spins individually (i.e., not in pairs) after time delays chosen randomly from the interval $[0,2]$. The results indicate that the heating becomes relatively weaker with increasing system sizes, which is in agreement with the general expectations from our analytical considerations.

\section{Final remarks and Conclusions} \label{sec_conclude}
Let us now make two final remarks.

(1) According to our stability criterion, ensembles with broad $g(E)$ are not stable. In particular, the QMC ensemble~\cite{qmc} is not stable with respect to local measurements, because it implies $g(E)$ having the form of two widely separated peaks corresponding to $T=0$ and $T=\infty$, respectively~\cite{qmc}.

(2) Our stability analysis can also be applied to unconventional quantum ensembles in experiments with nonmacroscopic isolated quantum clusters~\cite{blatt,hack,jelezko,jochim,greiner,newton_cradle,kai,shimizu}. Adapting the obtained results to such systems implies that the above experiments should avoid (i) external magnetic fields,  (ii) long-range order, and (iii) local constants of motion.

In conclusion, we have shown that even relatively rare local measurements impose strict constraints on quantum statistical ensembles. We introduced a stability criterion, according to which quantum statistical ensembles characterised by a total energy distribution significantly broader than that of a canonical ensemble are unstable. This result justifies the use of statistical ensembles with narrow $g(E)$ for equilibrium description of macroscopic systems.

\textit{Acknowledgements} - W.H. is grateful for the support from \textit{Studienstiftung des deutschen Volkes}. The authors acknowledge support by the state of Baden-Württemberg through bwHPC.

\appendix
\section{Width of $g(E)$ for the canonical ensemble} \label{app_width}
For the canonical ensemble,
\begin{equation} \label{exp}
 g(E)=\frac{1}{Z}e^{-\beta E}\nu(E)=\frac{1}{Z}e^{-\beta E+S(E)},
\end{equation}
where $Z=\int_{-\infty}^\infty e^{-\beta E}\nu(E)dE$ is the partition function, $\beta\equiv\frac{1}{T}$ ($k_B=1$) is the inverse temperature, and \mbox{$S(E)\equiv\ln\left[\frac{\nu(E)}{\nu_0}\right]$} is the entropy ($\nu_0$ is an unimportant constant).

The distribution $g(E)$ can be approximated by a Gaussian function $\exp\left(-\frac{(E-E_\text{av})^2}{2w_g^2}\right)$, where $E_\text{av}$ is defined by the condition $\beta=S'(E_\text{av})$ with $S'(E)\equiv\frac{dS(E)}{dE}$. Expanding the exponent in Eq.~\eqref{exp} at $E_\text{av}$ up to the second order, we obtain
\begin{eqnarray}
 w_g=\sqrt{-\frac{1}{\frac{d^2S(E)}{dE^2}\big|_{E_\text{av}}}}&=&T(E_\text{av})\sqrt{\frac{dE(T)}{dT}\bigg|_{T(E_\text{av})}}\\
 &=&T(E_\text{av})\sqrt{C_V(T(E_\text{av}))},\nonumber
\end{eqnarray}
where $C_V(T)\equiv\frac{dE(T)}{dT}$ is the total specific heat of the system, $T(E_\text{av})=\frac{1}{\beta(E_\text{av})}$, and $\beta(E)$ is the inverse function of $E(\beta)$ defined above. Typically, $C_V\sim N_s$. Therefore, $\frac{w_g}{E_{\text{av}}-E_{\min}}\sim\frac{1}{\sqrt{N_s}}$, which means that $g(E)$ is sharply peaked.

\section{Derivation of Eq.~\eqref{eqn_gen}} \label{app_derivation_5}
We denote the initial density matrix of the macroscopic system as $\rho_0$. We choose $\rho_0$ to be diagonal in the basis of the energy eigenstates $|E_i\rangle$,
\begin{equation} \label{eqn_rhoo}
 \rho_0=\sum_{i}(\rho_0)_{ii}|E_i\rangle\langle E_i|.
\end{equation}
By doing this, we assume that the off-diagonal elements of $\rho_0$, even if initially present, would have a negligible effect on the subsequent evolution of $g(E)$ due to the rapid dephasing between states having macroscopically different values of energy.

The transformation from $\rho_{n-1}$ to $\rho_n$ given in Eq.~\eqref{eqn_rho_next} can be iterated to obtain the transformation from $\rho_0$ to $\rho_n$,
\begin{equation} \label{sum}
 (\rho_n)_{kl}=\frac{1}{B}\langle E_k|{\cal O}\rho_0{\cal O}^\dagger|E_l\rangle,
\end{equation}
where $(\rho_n)_{kl}=\langle E_k|\rho_n|E_l\rangle$, $B$ is a normalisation factor, and
\begin{equation} \label{b}
 {\cal O}={\cal P}_{n}e^{-i{\cal H}(t_{n}-t_{n-1})}{\cal P}_{n-1}\cdots{\cal P}_{2}e^{-i{\cal H}(t_{2}-t_{1})}{\cal P}_{1}e^{-i{\cal H}t_{1}}.
\end{equation}
We are interested in the total-energy distribution and, therefore, focus on $(\rho_n)_{kk}$, i.e., the diagonal elements of the density matrix in the total-energy basis. Substituting Eq.~\eqref{b} and Eq.~\eqref{eqn_rhoo} into Eq.~\eqref{sum}, we obtain after a transformation
\begin{eqnarray}
 (\rho_n)_{kk}&=&\frac{1}{B}\langle E_k|{\cal O}\left(\sum_{i}(\rho_0)_{ii}|E_i\rangle\langle E_i|\right){\cal O}^\dagger|E_k\rangle\\
 &=&\frac{1}{B}(\rho_0)_{kk}\langle E_k|{\cal O}|E_k\rangle\langle E_k|{\cal O}^\dagger|E_k\rangle \label{first}\\
 &&+\frac{1}{B}\langle E_k|{\cal O}\left(\sum_{i,i\neq k}(\rho_0)_{ii}|E_i\rangle\langle E_i|\right){\cal O}^\dagger|E_k\rangle.\nonumber
\end{eqnarray}
Using the identity \mbox{$|E_k\rangle\langle E_k|=\mathbf{1}-\sum_{i,i\neq k}|E_i\rangle\langle E_i|$}, we further rewrite Eq.~\eqref{first} as
\begin{eqnarray}
 (\rho_n)_{kk}&=&\frac{1}{B}(\rho_0)_{kk}\langle E_k|{\cal O}^\dagger{\cal O}|E_k\rangle \label{eqn_trans}\\
 &&-\frac{1}{B}\sum_{i,i\neq k}(\rho_0)_{kk}\langle E_k|{\cal O}^\dagger|E_i\rangle\langle E_i|{\cal O}|E_k\rangle \nonumber\\
 &&+\frac{1}{B}\sum_{i,i\neq k}(\rho_0)_{ii}\langle E_i|{\cal O}^\dagger|E_k\rangle\langle E_k|{\cal O}|E_i\rangle. \nonumber
\end{eqnarray}
Now we introduce the coarse graining of the energy axis, which means that we divide the energy axis into bins of width $\Delta_\text{e}$ introduced above. Substituting Eq.~\eqref{eqn_trans} into Eq.~\eqref{eqn_defg}, we obtain
\begin{eqnarray} \label{eqn_cancel}
 g_n(E)&=&\frac{1}{\Delta_\text{e}B}\sum_k^{\text{bin}(E)}(\rho_0)_{kk}\langle E_k|{\cal O}^\dagger{\cal O}|E_k\rangle \\
 &&-\frac{1}{\Delta_\text{e}B}\sum_k^{\text{bin}(E)}\sum_{i,i\neq k}(\rho_0)_{kk}\langle E_k|{\cal O}^\dagger|E_i\rangle\langle E_i|{\cal O}|E_k\rangle\nonumber\\
 &&+\frac{1}{\Delta_\text{e}B}\sum_k^{\text{bin}(E)}\sum_{i,i\neq k}(\rho_0)_{ii}\langle E_i|{\cal O}^\dagger|E_k\rangle\langle E_k|{\cal O}|E_i\rangle, \nonumber
\end{eqnarray}
where $\sum$ [without $\text{bin}(E)$] denotes a sum which is not restricted to the bin.

\begin{figure*}[tb]
	\centering
	\includegraphics[width=13cm]{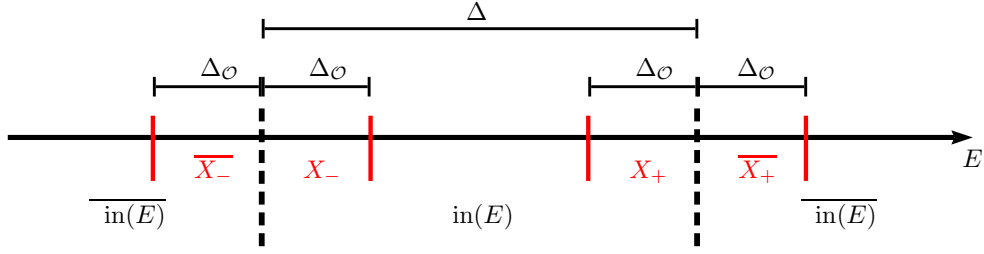}
	\caption{(Colour online) Schematic representation of the energy intervals $\text{bin}(E)$, $\overline{\text{bin}(E)}$, $X_-$, $\overline{X}_-$, $X_+$, and $\overline{X}_+$, introduced in the text. The characteristic size of the intervals $\Delta_\text{e}$ and $\Delta_{\cal O}$ are indicated above.}
	\label{fig_els}
\end{figure*}

Now we show that the first term in Eq.~\eqref{eqn_cancel} makes the main contribution to $g_n(E)$, while the last two terms in Eq.~\eqref{eqn_cancel} mainly compensate each other. To show this, we split each unrestricted sum into two sums, where one sum extends over the energy eigenstates within the bin, while the other one extends over the energy eigenstates outside the bin. The latter sum is to be denoted as $\sum^{\overline{\text{bin}(E)}}$. We also use the relation \mbox{$\langle E_k|{\cal O}^\dagger|E_i\rangle\langle E_i|{\cal O}|E_k\rangle=|\langle E_i|{\cal O}|E_k\rangle|^2$}. Hence, we obtain for the last two terms in Eq.~\eqref{eqn_cancel}
\begin{eqnarray} \label{eqn_nobin}
 -\frac{1}{\Delta_\text{e}B}\sum_k^{\text{bin}(E)}\sum^{\text{bin}(E)}_{i,i\neq k}(\rho_0)_{kk}|\langle E_i|{\cal O}|E_k\rangle|^2\\
 -\frac{1}{\Delta_\text{e}B}\sum_k^{\text{bin}(E)}\sum^{\overline{\text{bin}(E)}}_{i}(\rho_0)_{kk}|\langle E_i|{\cal O}|E_k\rangle|^2  \nonumber\\
 +\frac{1}{\Delta_\text{e}B}\sum_k^{\text{bin}(E)}\sum^{\text{bin}(E)}_{i,i\neq k}(\rho_0)_{ii}|\langle E_k|{\cal O}|E_i\rangle|^2 \nonumber\\
 +\frac{1}{\Delta_\text{e}B}\sum_k^{\text{bin}(E)}\sum^{\overline{\text{bin}(E)}}_{i}(\rho_0)_{ii}|\langle E_k|{\cal O}|E_i\rangle|^2.\nonumber
\end{eqnarray}
Now the two terms in Eq.~\eqref{eqn_nobin} where both sums are restricted to bin$(E)$ cancel each other. This can be readily seen after exchanging the summation indices $i$ and $k$ in one of these two terms. The remaining terms in Eq.~\eqref{eqn_cancel} can be further rewritten as
\begin{eqnarray} \label{eqn_compare}
 g_n(E)&=&\frac{1}{\Delta_\text{e}B}\sum_k^{\text{bin}(E)}(\rho_0)_{kk}\langle E_k|{\cal O}^\dagger{\cal O}|E_k\rangle\\
 &&-\frac{1}{\Delta_\text{e}B}\sum_k^{\text{bin}(E)}(\rho_0)_{kk}\left(\sum^{\overline{\text{bin}(E)}}_{i}|\langle E_i|{\cal O}|E_k\rangle|^2\right)\nonumber\\
 &&+\frac{1}{\Delta_\text{e}B}\sum^{\overline{\text{bin}(E)}}_{i}(\rho_0)_{ii}\left(\sum_k^{\text{bin}(E)}|\langle E_k|{\cal O}|E_i\rangle|^2\right). \nonumber
\end{eqnarray}
The last two terms of Eq.~\eqref{eqn_compare} contain off-diagonal elements $|\langle E_i|{\cal O}|E_k\rangle|^2$ corresponding to transitions between energy bins, because $E_k$ and $E_i$ lie in different bins. Let us denote the characteristic energy range $|E_k - E_i|$ of the off-diagonal elements \mbox{$\langle E_k|{\cal O}|E_i\rangle$} as $\Delta_{\cal O}$. This range is limited by the condition
\begin{equation}
 \Delta_{\cal O} \sim \epsilon_1 n \ll \Delta_\text{e}.
\end{equation}
Therefore, only small energy intervals of length $\Delta_{\cal O}$ near the boundaries between the bins contribute to the sums. As shown in Fig.~\ref{fig_els}, we label these energy intervals as $X_-$, $\overline{X}_-$,  $X_+$ and $\overline{X}_+$. With such notations, Eq.~\eqref{eqn_compare} can be rewritten as 
\begin{eqnarray} \label{eqn_cut}
 g_n(E)&=&\frac{1}{\Delta_\text{e}B}\sum_k^{\text{bin}(E)}(\rho_0)_{kk}\langle E_k|{\cal O}^\dagger{\cal O}|E_k\rangle\\
 &&-\frac{1}{\Delta_\text{e}B}\sum_k^{X_-}(\rho_0)_{kk}\left(\sum^{\overline{X_-}}_{i}|\langle E_i|{\cal O}|E_k\rangle|^2\right)\nonumber\\
 &&-\frac{1}{\Delta_\text{e}B}\sum_k^{X_+}(\rho_0)_{kk}\left(\sum^{\overline{X_+}}_{i}|\langle E_i|{\cal O}|E_k\rangle|^2\right)\nonumber\\
 &&+\frac{1}{\Delta_\text{e}B}\sum^{\overline{X_-}}_{i}(\rho_0)_{ii}\left(\sum_k^{X_-}|\langle E_k|{\cal O}|E_i\rangle|^2\right)\nonumber\\
 &&+\frac{1}{\Delta_\text{e}B}\sum^{\overline{X_+}}_{i}(\rho_0)_{ii}\left(\sum_k^{X_+}|\langle E_k|{\cal O}|E_i\rangle|^2\right).\nonumber
\end{eqnarray}

Now we show that the last four terms in Eq.~\eqref{eqn_cut} can be neglected in comparison with the first one, provided $g_0(E)$ does not change too fast. Specifically, we impose the condition~\eqref{eqn_ineq}, which together with the inequalities~\eqref{bins} yields
\begin{equation} \label{eqn_cond}
 \left|\frac{dg_0(E)}{dE}\right|\lesssim \frac{g_0(E)}{w_\text{can}}\ll \frac{g_0(E)}{\Delta_\text{e}},
\end{equation}
where $w_\text{can}= T(E_\text{av}) \sqrt{C_V(E_\text{av})}$ is the width of the energy distribution corresponding to the canonical ensemble with the same initial average energy as that of $g_0(E)$. The condition in Eq.~\eqref{eqn_cond} must be satisfied within the energy interval, where $g_0(E)$ is large enough to make a non-negligible contribution to the normalisation integral $\int_{-\infty}^\infty g_0(E) dE=1$.

In the first term of Eq.~\eqref{eqn_cut}, we expect that, even if $(\rho_0)_{kk}$ and $\langle E_k|{\cal O}^\dagger{\cal O}|E_k\rangle$ fluctuate with respect to their bin-averaged values, they do it in an uncorrelated way.  According to Eq.~\eqref{eqn_defg}, the bin-averaged value of $(\rho_0)_{kk}$ is $\Delta_\text{e} g_0(E)/N_\text{bin}(E)$, where $N_\text{bin}(E)$  is the number of states within the bin. We define the bin average of $\langle E_k|{\cal O}^\dagger{\cal O}|E_k\rangle$ as
\begin{equation} \label{eqn_bin_def}
 \left[{\cal O}^\dagger{\cal O}\right]_{\text{diag}}\!(E)\equiv\frac{1}{N_\text{bin}(E)}\sum_k^{\text{bin}(E)}\langle E_k|{\cal O}^\dagger{\cal O}|E_k\rangle.
\end{equation}
Given the right inequalities in Eqs.~\eqref{eqn_defg} and \eqref{eqn_cond}, both bin averages change very weakly over the bin size $\Delta_\text{e}$. Therefore, we can approximate the entire first term in Eq.~\eqref{eqn_cut} as
\begin{equation} \label{eqn_1st_term}
 \frac{1}{\Delta_\text{e}B}\!\sum_k^{\text{bin}(E)}(\rho_0)_{kk}\langle E_k|{\cal O}^\dagger{\cal O}|E_k\rangle
 \!\approx\!
 \frac{1}{B}\left[{\cal O}^\dagger{\cal O}\right]_{\text{diag}}\!(E)\ g_0(E).
\end{equation}
Each of the remaining four terms in Eq.~\eqref{eqn_cut} has comparable values. Let us estimate the first of them. We use the following inequality
\begin{equation} \label{eqn_leq}
 \sum^{\overline{X_-}}_{i}|\langle E_i|{\cal O}|E_k\rangle|^2\leq\sum_{i}|\langle E_i|{\cal O}|E_k\rangle|^2=\langle E_k|{\cal O}^\dagger{\cal O}|E_k\rangle,
\end{equation}
where, as before, the second sum extends over all energy eigenstates of the system. Employing this inequality together with the assumptions used for deriving Eq.~\eqref{eqn_1st_term}, we obtain
\begin{eqnarray} \label{eqn_startiq}
 &&\frac{1}{\Delta_\text{e}B}  \sum^{X_-}_{k}(\rho_0)_{kk}  \left( \sum_i^{\overline{X}_-}|\langle E_i|{\cal O}|E_k\rangle|^2 \right)\\
 &&\leq  \frac{1}{\Delta_\text{e}B}  \sum^{X_-}_{k}(\rho_0)_{kk} \langle E_k|{\cal O}^\dagger{\cal O}|E_k\rangle \nonumber\\
 &&\approx \frac{1}{\Delta_\text{e}B}\sum^{X_-}_{k}(\rho_0)_{kk}\left[{\cal O}^\dagger{\cal O}\right]_{\text{diag}}\!(E) \nonumber\\
 &&\approx \frac{1}{B} \left[{\cal O}^\dagger{\cal O}\right]_{\text{diag}}\!(E)\frac{\Delta_{\cal O}}{\Delta_\text{e}}g_0(E). \nonumber
\end{eqnarray}
Since $\frac{\Delta_{\cal O}}{\Delta_\text{e}} \ll 1$, Eqs.~\eqref{eqn_cut}, \eqref{eqn_1st_term} and \eqref{eqn_startiq} imply that
\begin{equation} \label{eqn_av_app_2}
 g_n(E) \approx \frac{1}{B}\left[{\cal O}^\dagger{\cal O}\right]_{\text{diag}}\!(E)\ g_0(E),
\end{equation}
which is the same as Eq.~\eqref{eqn_gen}. 

\section{Derivation of Eq.~\eqref{eqn_av_app}} \label{app_derivation_6}
Let us now consider the cutting function in Eq.~\eqref{eqn_av_app_2} or, equivalently, in Eq.~\eqref{eqn_gen}
\begin{eqnarray} 
 &&\left[{\cal O}^\dagger{\cal O}\right]_{\text{diag}}\!(E)\\
 &&=\!\!\Big[{\cal P}^{\dagger}_1e^{i{\cal H}(t_{2}-t_{1})}{\cal P}^{\dagger}_{2}\!\cdots\!{\cal P}^{\dagger}_{n}{\cal P}_{n}\!\cdots\!{\cal P}_{2}e^{-i{\cal H}(t_{2}-t_{1})}{\cal P}_1 \Big]_{\text{diag}}\!(E)\nonumber\\
 &&=\!\!\Big[{\cal P}^{\dagger}_1(t_1){\cal P}^{\dagger}_{2}(t_2)\!\cdots\!{\cal P}^{\dagger}_{n}(t_n){\cal P}_{n}(t_n)\!\cdots\!{\cal P}_{2}(t_2){\cal P}_1(t_1) \Big]_{\text{diag}}\!(E), \nonumber
\end{eqnarray}
where, in the last expression, we used \mbox{${\cal P}^{\dagger}_k(t_k)\equiv e^{i{\cal H}t_{k}}{\cal P}^{\dagger}_k e^{-i{\cal H}t_{k}}$}, cf. Eq.~\eqref{eqn_bin_def}. Given that $n\ll\sqrt{N_s}$, it is rather unlikely that among the measured spins $m_1$, $m_2$, ..., $m_n$ there are two spins which are close to each other. Therefore, we assume the corresponding measurement operators ${\cal P}^{\dagger}_k(t_k)$ to commute. Rearranging the operators in the above expression and using ${\cal P}^{\dagger}_k(t_k){\cal P}^{\dagger}_k(t_k)={\cal P}^{\dagger}_k(t_k)$, we obtain
\begin{equation}
 \left[{\cal O}^\dagger{\cal O}\right]_{\text{diag}}\!(E)=\Big[{\cal P}_{n}(t_n)\cdots{\cal P}_{2}(t_2){\cal P}_1(t_1) \Big]_{\text{diag}}\!(E).
\end{equation}
On the right-hand side of the above expression, there is the $n$-particle correlation function for the microcanonical ensemble. For systems with short-range interaction, we expect that correlation functions of mutually independent quantities factorise, which leads to
\begin{eqnarray} \label{eqn_factorise}
 &&\Big[{\cal P}_{n}(t_n)\cdots{\cal P}_{2}(t_2){\cal P}_1(t_1) \Big]_{\text{diag}}\!(E)\\
 &&=\Big[{\cal P}_{n}(t_n)\Big]_{\text{diag}}\!(E)\cdots\Big[{\cal P}_{2}(t_2)\Big]_{\text{diag}}\!(E)\Big[{\cal P}_1(t_1) \Big]_{\text{diag}}\!(E).\nonumber
\end{eqnarray}
This allow us to treat measurements iteratively. Since $g(E)$ does not change with time between the measurements, the result does not depend on the particular values of $t_1$, ..., $t_n$ and, according to Eq.~\eqref{eqn_bin_def}, \mbox{${\cal P}_{k}(t_k)={\cal P}_{k}(0)={\cal P}_{k}$}. Therefore, we obtain
\begin{equation} \label{this}
 g_n(E)=\frac{1}{B_n} \left[  {\cal P}_n\right]_{\text{diag}}\!(E)\  g_{n-1}(E),
\end{equation}
which is the same as Eq.~\eqref{eqn_av_app}.

For macroscopic systems, the above analysis does not depend on the particular value of $\tau_\text{m}$. However, for small clusters, Eq.~\eqref{this} is valid as long as the delay between measurements $\tau_\text{m}$ is much longer than the characteristic time scales of the system's microscopic dynamics. These characteristic time scales are, for example, of the order of $\frac{1}{H_z}$ for Hamiltonian~\eqref{eqn_ham1} or of the order of \mbox{$\frac{1}{J}=\frac{1}{\sqrt{J_x^2+J_y^2+J_z^2}}$} for Hamiltonian~\eqref{eqn_ham2}.

There may be anomalous situations, when the expression~\eqref{eqn_gen} for the measurement of a pair of neighbouring spins cannot be approximated by the result of two successive applications of Eq.~\eqref{this} no matter how long the delay between the two measurements is. An example is the Ising model defined by $J_x = J_y = 0$ in Hamiltonian~\eqref{eqn_ham2} which has anomalously many local integrals of motion ($z$ components of spins).

\section{Derivation of Eq.~\eqref{eqn_enmod_h}} \label{app_derivation_9}
The Hamiltonian ${\cal H}=-H_z\sum_iS_{iz}$ with $H_z>0$ for spins in a magnetic field is diagonal in the product basis of individual spin states $|\!\!\uparrow\rangle_z$ and $|\!\!\downarrow\rangle_z$. Therefore, we obtain for the expectation value of the $z$ polarisation of spin at lattice site $m_n$,
\begin{equation}
 \langle S_{nz} \rangle= \frac{\langle(S_z)_{\text{tot}}\rangle}{N_s}=-\frac{E}{N_sH_z}=-\frac{E}{E_{\max}-E_{\min}}.
\end{equation}
With $\langle S_n\rangle=\langle S_{nz}\rangle\cos(\vartheta_n)$, this yields
\begin{equation} \label{eqn_spinmag}
 \left[S_n\right]_{\text{diag}}\!({E})=-\cos(\vartheta_n)\frac{E}{E_{\max}-E_{\min}},
\end{equation}
which, taking into account the relation ${\cal P}_n=\frac{1}{2}\mathbf{1}+S_n$, results in Eq.~\eqref{eqn_enmod_h}.

\section{Calculation of the results shown in Figure~\ref{fig_stab}} \label{app_fig}
Substituting $g_0(E)=\frac{1}{2}[\delta(E-E_1)+\delta(E-E_2)]$ into Eq.~\eqref{eqn_dg_av}, we obtain
\begin{eqnarray}
 \overline{\Delta G}(n)&=&\frac{1}{2}\overline{\bigg|\prod_{i=1}^n\frac{1}{B_i}\left[ {\cal P}_i \right]_{\text{diag}}\!(E_1) - 1\bigg|}\\
 &&+\frac{1}{2}\overline{\bigg|\prod_{i=1}^n\frac{1}{B_i}\left[ {\cal P}_i \right]_{\text{diag}}\!(E_2) - 1\bigg|},\nonumber
\end{eqnarray}
where the bar denotes the result of averaging. The probability for obtaining the given measurement outcomes equals the normalisation factor $\prod_{i=1}^n B_i$. Given Eq.~\eqref{eqn_enmod_h}, the above expression only depends on the polar angles $\vartheta_1$, ..., $\vartheta_n$ describing the measurement outcome. This leads to
\begin{widetext}
\begin{eqnarray}
 \overline{\Delta G}(n)\!&=&\frac{1}{2}\left[\int_0^\pi\prod_{i=1}^n B_i\bigg|\prod_{i=1}^n\frac{1}{B_i}\left[ {\cal P}_i \right]_{\text{diag}}\!(E_1) - 1\bigg|\sin(\vartheta_1)\cdots\sin(\vartheta_n)\ d\vartheta_1\cdots d\vartheta_n\right.\nonumber\\
 &&\left.+\int_0^\pi\prod_{i=1}^n B_i\bigg|\prod_{i=1}^n\frac{1}{B_i}\left[ {\cal P}_i \right]_{\text{diag}}\!(E_2) - 1\bigg|\sin(\vartheta_1)\cdots\sin(\vartheta_n)\ d\vartheta_1\cdots d\vartheta_n\right]\\
 &=&\frac{1}{2}\left[\int_0^\pi\bigg|\prod_{i=1}^n\left[ {\cal P}_i \right]_{\text{diag}}\!(E_1) - \prod_{i=1}^n B_i\bigg|\sin(\vartheta_1)\cdots\sin(\vartheta_n)\ d\vartheta_1\cdots d\vartheta_n\right.\nonumber\\
 &&\left.+\int_0^\pi\bigg|\prod_{i=1}^n\left[ {\cal P}_i \right]_{\text{diag}}\!(E_2) - \prod_{i=1}^n B_i\bigg|\sin(\vartheta_1)\cdots\sin(\vartheta_n)\ d\vartheta_1\cdots d\vartheta_n\right].\label{eqn_here}
\end{eqnarray}
\end{widetext}
Using Eq.~\eqref{eqn_enmod_h}, we further obtain
\begin{eqnarray}
 \prod_{i=1}^n B_i&=&\int\prod_{i=1}^n\left[ {\cal P}_i \right]_{\text{diag}}\!(E)g_0(E)\ dE\\
 &=&\frac{1}{2}\prod_{i=1}^n \left(\frac{1}{2}-\cos(\vartheta_i)\frac{E_1}{E_{\max}-E_{\min}}\right)\nonumber\\
 &&+\frac{1}{2}\prod_{i=1}^n \left(\frac{1}{2}-\cos(\vartheta_i)\frac{E_2}{E_{\max}-E_{\min}}\right).\nonumber
\end{eqnarray}
Substituting the last expression into Eq.~\eqref{eqn_here}, the two summands in Eq.~\eqref{eqn_here} are identical and this leads to
\begin{eqnarray}
 \overline{\Delta G}(n)\!=\frac{1}{2}\int_0^\pi\Bigg[\bigg|\prod_{i=1}^n \left(\frac{1}{2}-\cos(\vartheta_i)\frac{E_1}{E_{\max}-E_{\min}}\right)\\
 -\prod_{i=1}^n \left(\frac{1}{2}-\cos(\vartheta_i)\frac{E_2}{E_{\max}-E_{\min}}\right)\bigg|\Bigg]\nonumber\\
 \sin(\vartheta_1)\cdots\sin(\vartheta_n)\ d\vartheta_1\cdots d\vartheta_n.\nonumber
\end{eqnarray}
Integrating the above expression numerically, we obtain the results shown in Fig.~\ref{fig_stab}.

\section{Derivation of the approximation for $\overline{\Delta G}(n)$ plotted in Figure~\ref{fig_stab}} \label{app_anexp}
In this section, we derive the analytical approximation for $\overline{\Delta G}(n)$ used in Fig.~2. From relation~\eqref{this}, we obtain 
\begin{equation} \label{mult}
 g_n(E)=\prod_{i=1}^{n}\frac{1}{B_i}\left[{\cal P}_i\right]_{\text{diag}}\!(E)\ g_0(E).
\end{equation}
The substitution of this expression into Eq.~\eqref{DG} leads to
\begin{equation} \label{eqn_integral}
 \Delta G(n)=\int_{- \infty}^{+ \infty} \Big| \prod_{i=1}^{n}\frac{1}{B_i}\left[{\cal P}_i\right]_{\text{diag}}\!(E)-1\Big| g_0(E) dE,
\end{equation}
which, after averaging, gives Eq.~\eqref{eqn_dg_av}.

For the initial distribution $g_0(E)=\frac{1}{2}[\delta(E-E_1)+\delta(E-E_2)]$, Eq.~\eqref{mult} leads to \linebreak$g_n(E)=p_1\delta(E-E_1)+p_2\delta(E-E_2)$ with some probabilities $p_1$ and $p_2$ ($p_1+p_2=1$). We further note that for $n\to\infty$, either $p_1\to0$ or $p_2\to0$, such that $\Delta G(n)=|p_1-p_2|\to1$.

In order to find an approximate expression for $\overline{\Delta G}(n)$, let us first observe that
\begin{equation}
 \overline{\Delta G^2}(n)\approx\overline{[\Delta G(n)-\Delta G(n-1)]^2}+\overline{\Delta G^2(n-1)},
\end{equation}
because $\overline{\Delta G(n)-\Delta G(n-1)}\approx0$. Therefore, we make the ansatz $\overline{\Delta G^2}(n)=\sum_{i=1}^n \overline{\gamma_i^2}$, where $\gamma_i=\Delta G(i)-\Delta G(i-1)$. Since the individual summands $\overline{\gamma_i^2}$ become smaller as $\overline{\Delta G^2}(n)$ approaches 1, we make the rough approximation $\overline{\gamma_i^2}\cong 1-\overline{\Delta G^2}(n)$, which, in the continuum limit for $n$, leads to the differential equation $\frac{d\overline{\Delta G^2}(n)}{dn}=\lambda[1-\overline{\Delta G^2}(n)]$, where $\lambda$ is some constant. Assuming that $\overline{\Delta G^2}(n)\approx\overline{\Delta G}^2(n)$, we obtain
\begin{equation} \label{rees}
 \overline{\Delta G}(n)\approx\sqrt{1-e^{-\lambda n}}.
\end{equation}
We further adopt an approximation
\begin{equation} \label{eqn_lam}
 \lambda = \kappa\ u^2 \left(E_1-E_2\right)^2,
\end{equation}
where $u \equiv  \overline{|d\left[{\cal P}_n\right]_{\text{diag}}\!(E)/dE|}$, and $\kappa$ is a fitting parameter. The expression in Eq.~\eqref{eqn_lam} can be justified by the following considerations. The parameter $\lambda$ must be equal to 0 when $E_1=E_2$. Also, $\lambda$ must remain invariant under a sign change of $E_1-E_2$. Therefore, the lowest-order term allowed in an analytical expansion of $\lambda$ around 0 is proportional to $\left(E_1-E_2\right)^2$. Assuming that the value of $\lambda$ is only controlled by $u$ and $E_1-E_2$, we conclude that $u$ must also enter quadratically in order for $\lambda$ to be dimensionless.

In Fig.~2, all three curves have been plotted with the same value for the parameter $\kappa$ such that $\lambda = 0.3\ \left(\frac{E_1-E_2}{E_{\max}-E_{\min}}\right)^2$.

The above approximation can be further supported by a more detailed calculation of $\overline{\Delta G}(n)$ in which case the expression for $\Delta G(n)$ in Eq.~\eqref{eqn_integral} is to be averaged over all possible measurement outcomes. For this average, $\Delta G(n)$ must be weighed by the probability for a given set of $n$ measurement outcomes, which equals the normalisation coefficient $\prod_{i=1}^{n}{N_i}=\int \prod_{i=1}^{n}\left[{\cal P}_i\right]_{\text{diag}}\!(E)\ g_0(E)dE$. Let us now make a simplifying assumption that the spin measurements are only done along the $z$ direction. Consequently, there are two possible measurement outcomes: positive ($\vartheta_n=0$) and negative ($\vartheta_n=\pi$). According to Eq.~\eqref{eqn_spinmag}, the projection operator is thus $\left[{\cal P}_i\right]_{\text{diag}}\!(E)=\frac{1}{2}\pm\frac{E}{E_{\max}-E_{\min}}$. After averaging over these two possibilities for each measurement, we obtain
\begin{equation} \label{binomial}
 \overline{\Delta G}(n)=\frac{1}{2}\sum_{k=0}^{n}\left| D_{np_1}(k)-D_{np_2}(k)\right|,
\end{equation} 
where $D_{np}(k)=\binom{n}{k}p^k(1-p)^{n-k}$ is the binomial distribution and $p_i=\frac{1}{2}-\frac{E_i}{E_{\max}-E_{\min}}$. The value of $\overline{\Delta G}(n)$ is governed by the overlap as function of $k$ between the two binomial distributions $D_{np_1}(k)$ and $D_{np_2}(k)$. With an increasing $n$, this overlap decreases approximately exponentially, which is consistent with the asymptotic behaviour of $\overline{\Delta G}(n)$ that follows from Eq.~\eqref{rees} for large $n$.

\section{Narrowing of a Gaussian probability distribution} \label{app_gauss}
For a Gaussian distribution
\begin{equation}
 g_{n-1}(E)\cong \exp[-\frac{(E-E_\text{av})^2}{2w^2_{g,n-1}}]
\end{equation}
with the variance $w^2_{g,n-1}\ll(E_{\max}-E_{\min})^2$, the ``cutting'' by the linear function $\left[{\cal P}_n\right]_{\text{diag}}\!(E)$ can be expressed as
\begin{eqnarray}
g_n(E)\sim \left[{\cal P}_n\right]_{\text{diag}}\!(E)\ e^{-\frac{(E-E_\text{av})^2}{2w^2_{g,n-1}}}\\
=e^{\ln(\left[{\cal P}_n\right]_{\text{diag}}\!(E))-\frac{(E-E_\text{av})^2}{2w^2_{g,n-1}}}.\nonumber
\end{eqnarray}
It changes the width according to the relation
\begin{equation} \label{wid}
 \frac{1}{w_{g,n}^2}-\frac{1}{w_{g,n-1}^2}=\left(\frac{\left[{\cal P}_n\right]'_{\text{diag}}(E_\text{av})}{\left[{\cal P}_n\right]_{\text{diag}}(E_\text{av})}\right)^2,
\end{equation}
where $\left[{\cal P}_n\right]'_{\text{diag}}\!(E_\text{av})=\frac{d\left[{\cal P}_n\right]_{\text{diag}}\!(E_\text{av})}{dE_\text{av}}$. From Eq.~\eqref{wid}, we obtain
\begin{equation} \label{widg}
 \overline{\left[\frac{1}{w_{g,n}^2}\right]}\approx \frac{1}{w_{g,0}^2}+u^2\ n,
\end{equation}
where $u=\overline{|\left[{\cal P}_n\right]'_{\text{diag}}(E_\text{av})|}$, and we use $\left[{\cal P}_n\right]_{\text{diag}}(E_\text{av})\sim1$.

\section{Derivation of Eq.~\eqref{eqn_modi}} \label{app_derivation_13}
If two measurements $n-1$ and $n$ accidentally occur close in space and time, the corresponding projection operators ${\cal P}_{n-1}(t_{n-1})$ and ${\cal P}_n(t_n)$ normally do not commute, and the effect of the two measurements does not factorise. Assuming that all other measured spins are far away from $m_{n-1}$ and $m_n$, we obtain for the term describing the effect of the two measurements entering the expression in Eq.~\eqref{eqn_factorise} $\left[{\cal A}_{n,n-1}^\dagger{\cal A}_{n,n-1}\right]_{\text{diag}}\!(E)$, where ${\cal A}_{n,n-1}={\cal P}_ne^{-i{\cal H}(t_{n}-t_{n-1})}{\cal P}_{n-1}$. The expression for the modification of $g(E)$ analogous to Eq.~\eqref{this} reads
\begin{equation} \label{rew}
 g_n(E)=\frac{1}{B_n}\left[{\cal A}_{n,n-1}^\dagger{\cal A}_{n,n-1}\right]_{\text{diag}}\!(E)\ g_{n-2}(E).
\end{equation}
For the function $\left[{\cal A}_{n,n-1}^\dagger{\cal A}_{n,n-1}\right]_{\text{diag}}\!(E)$, we obtain
\begin{eqnarray}
 &&\left[{\cal A}_{n,n-1}^\dagger{\cal A}_{n,n-1}\right]_{\text{diag}}\!(E)\\
 &&=\left[{\cal P}^\dagger_{n-1}(t_{n-1}){\cal P}^\dagger_{n}(t_{n}){\cal P}_{n}(t_{n}){\cal P}_{n-1}(t_{n-1})\right]_{\text{diag}}\!(E) \nonumber\\
 &&=\Big[{\cal P}_{n-1}(t_{n-1}){\cal P}_{n}(t_{n}){\cal P}_{n-1}(t_{n-1})\Big]_{\text{diag}}\!(E).\nonumber
\end{eqnarray}
Substituting Eq.~\eqref{eqn_sp} leads to Eq.~\eqref{eqn_modi}, which we rewrite here~\footnote{The bracket does not denote a commutator in the expression~\eqref{eqn_two_last}. For the definition of $[ \cdots ]_{\text{diag}}\!(E)$, we refer the reader to the discussion following Eq.~\eqref{eqn_gen}.},
\begin{eqnarray}\label{eqn_two_last}
 &&\left[{\cal A}_{n,n-1}^\dagger{\cal A}_{n,n-1}\right]_{\text{diag}}\!(E)\\
 &&=\frac{1}{4}+\frac{1}{2}\Big[\left\{S_{n-1}(t_{n-1}),S_{n}(t_{n}) \right\}\Big]_{\text{diag}}\!(E)\nonumber\\
 &&+\Big[S_{n-1}(t_{n-1})S_{n}(t_{n})S_{n-1}(t_{n-1})\Big]_{\text{diag}}\!(E),\nonumber
\end{eqnarray}
where $\left\{S_{n-1}(t_{n-1}),S_{n}(t_{n}) \right\}\equiv S_{n-1}(t_{n-1})S_{n}(t_{n})+S_{n}(t_{n})S_{n-1}(t_{n-1})$. When deriving Eq.~\eqref{eqn_two_last}, we assumed that the outcomes of single-spin measurements are not correlated with the total energy as explained in the main text. This is formally expressed as $\left[S_n\right]_{\text{diag}}\!(E)=0$. We further note that the three-spin term in Eq.~\eqref{eqn_two_last} should typically be significantly smaller than the two-spin term.

\section{Numerical investigation of spin systems} \label{app_num}
For the numerical investigation of interacting spin systems, we use the approach established in Refs.~\cite{dobrov,tarek_prl,raedt_fast}. This approach is based on the direct integration of the Schrödinger equation and, hence, does not involve the diagonalisation of the Hamiltonian matrix.

For the numerical study, we employ the property of quantum typicality, which means that, for quantum systems with very large Hilbert spaces, a single wave function representing a certain ensemble, for example, the ensemble given by $g(E)$, leads to the same average values of interest as the overall ensemble, cf. Eq.~\eqref{eqn_two_last}. Therefore, we study the stability of energy distributions by considering randomly chosen wave functions.

We prepare the two-peak $g_0(E)$ shown in Fig.~3 by creating a quantum superposition of two pure states each corresponding to a canonical ensemble, one with temperature $T_1=0.1$ and the other with temperature $T_2=-0.1$. In order to prepare the state at a temperature $T_1$ (or $T_2$), we use the method of imaginary-time evolution. First, we randomly select a wave function corresponding to the infinite-temperature limit~\cite{tarek_prl}. Second, we use the relation $e^{-\frac{{\cal H}}{T}}|\psi\rangle=e^{-i {\cal H}t}|\psi\rangle$, where $t=-i\frac{1}{T}$. For the imaginary time evolution, we use the same technique as for the one for the real-time evolution introduced in Refs.~\cite{tarek_prl,steinigeweg_1,steinigeweg_2} and also explained below.

For the real-time evolution, we use the fourth-order Runge-Kutta method~\cite{tarek_prl,steinigeweg_1,steinigeweg_2}
\begin{equation} \label{eqn_rk}
 |\psi(t+\Delta t)\rangle\approx |\psi(t)\rangle+|\nu_1\rangle+|\nu_2\rangle+|\nu_3\rangle+|\nu_4\rangle,
\end{equation}
where $|\nu_1\rangle=-i{\cal H}|\psi(t)\rangle\Delta t$, $|\nu_2\rangle=-\frac{1}{2}i{\cal H}|\nu_1\rangle\Delta t$, $|\nu_3\rangle=-\frac{1}{3}i{\cal H}|\nu_2\rangle\Delta t$, and $|\nu_4\rangle=-\frac{1}{4}i{\cal H}|\nu_3\rangle\Delta t$. For the Schrödinger equation, this method corresponds to the fourth-order Taylor expansion of the time-evolution operator. It was empirically shown in Ref.~\cite{tarek_prl} that this method is indeed accurate for the time intervals of interest in the present calculations.

In order to measure individual spins, we first choose a random direction $(\vartheta_n,\varphi_n)$ and then, using Eq.~\eqref{eqnrojector}, calculate ${\cal P}_n|\psi\rangle$.

For the calculation of the energy distribution $g_n(E)$ after the $n$th measurement at time $t_n$, we perform the following Fourier transform~\cite{raedt_fast}
\begin{equation}
 g_n(E)\cong\int \langle\psi(t+t_n)|\psi(t_n)\rangle\ e^{iEt}\ dt.
\end{equation}
In order to do this, we first generate a discrete time series by time evolving the wave function immediately after each measurement. In our calculations, the length of the time steps is 0.025 and the number of these time steps is 2048. After generating the time series, we multiply it by the Kaiser-Bessel window function in order to improve the energy resolution. The Kaiser-Bessel window function is defined as
\begin{equation}
  {\cal K}(k)\equiv \frac{I_0\left(\pi\alpha\sqrt{1-\left(\frac{2k}{N-1}-1\right)^2}\right)}{I_0(\pi\alpha)},
\end{equation}
where $I_0$ is the zeroth-order modified Bessel function of the first kind, $\alpha$ is a non-negative real number which determines the shape of the window, and $N=2048$. In our calculations, we use $\alpha=3$. Finally, we Fourier transform the resulting time series and normalise the distribution in order to obtain $g_n(E)$.

\end{document}